\documentclass[a4paper]{jpconf}
\usepackage{graphicx}
\usepackage[lcgreekalpha]{stix}	
\usepackage{siunitx}
    \DeclareSIUnit[]\c{\text{\emph{c}}}
    \DeclareSIUnit[per-mode = symbol]\MeVc{\MeV\per\c}
    \DeclareSIUnit\bit{\text{bit}}
    \DeclareSIUnit\Samples{\text{Samples}}
    \DeclareSIUnit\atp{\text{at}\%}

\usepackage{url}
\usepackage[sorting=none, backend=biber, style=chem-acs, dateabbrev=false, language=british, doi=true]{biblatex}
\usepackage{graphicx}
\usepackage{graphbox} 
\usepackage{adjustbox}
\usepackage{epstopdf}
\usepackage[font={small}]{caption}
\usepackage{subcaption}
\usepackage{tikz}														

\tikzset{font={\fontsize{8pt}{12}\selectfont}}							
\usepackage[final]{pdfpages}
\usepackage{chngcntr} 
\usepackage[bottom]{footmisc} 

\graphicspath{{figures}}

\usepackage{booktabs}
\usepackage{multirow}
\usepackage{rotating}

\usepackage{amsmath}

\DeclareRobustCommand\upgamma{\ensuremath{\mathrm\gamma}}
\DeclareRobustCommand\upmu{\ensuremath{\mathrm\mu}}
\DeclareRobustCommand\uppi{\ensuremath{\mathrm\pi}}
\DeclareSIUnit\bar{bar}

\newcommand*\ie{\emph{i.e.}\ }
\newcommand*\eg{\emph{e.g.}\ }
\newcommand*\musr{\ensuremath{\upmu\text{SR}}}
\newcommand*\LN{\ensuremath{\text{LN}_2}}
\newcommand{\email}[1]{\href{mailto:#1}{#1}}
\newcommand*\vs{\emph{vs.}\ }

\usepackage{csquotes}
\usepackage[acronym]{glossaries}

\newacronym{mexp}{GIANT}{GermanIum Array for Non-destructive Testing}
\newacronym{smus}{\ensuremath{\text{S}{\upmu}\text{S}}}{Swiss Muon Source}

\usepackage[breaklinks=true]{hyperref}
\usepackage[capitalise]{cleveref}
\usepackage{pdfpages} 
\hypersetup
{
    pdfauthor       = {L.\ Gerchow, S.\ Biswas, G.\ Janka, A.\ Amato},
    pdfsubject      = {MIXE},
    pdftitle        = {MIXE at PSI},
    pdfkeywords     = {MIXE, elemental analysis, muon, PSI},
    colorlinks      = true,
}

\begin{document}
\title{\Gls{mexp} setup for Muon Induced X-ray Emission (MIXE) at the Paul Scherrer Institute}
\author{Lars Gerchow$^{1,*}$, Sayani Biswas$^{1,*}$, Gianluca Janka$^1$, Carlos Vigo$^1$, Andreas Knecht$^2$, Stergiani Marina Vogiatzi$^{2,3}$, Narongrit Ritjoho$^4$, Thomas Prokscha$^1$, Hubertus Luetkens$^1$ and Alex Amato$^{1,*}$}

\address{$^1$ Laboratory of Muon Spin Spectroscopy, Neutron and Muon Division, Paul Scherrer Institute, Forschungsstrasse 111, 5232 PSI-Villigen, Switzerland}
\address{$^2$ Laboratory of Particle Physics, Neutron and Muon Division, Paul Scherrer Institute, Forschungsstrasse 111, 5232 PSI-Villigen, Switzerland}
\address{$^3$ Institute for Particle Physics and Astrophysics, ETH Zürich, 8093 Zürich, Switzerland}
\address{$^4$ School of Physics, Institute of Science, Suranaree University of Technology, 111 University Avenue Muang, Nakhon Ratchasima 30000, Thailand}

\ead{\email{lars.gerchow@psi.ch}, \email{sayani.biswas@psi.ch}, \email{gianluca.janka@psi.ch}, \email{carlos.vigo@psi.ch}, \email{andreas.knecht@psi.ch}, \email{stella.vogiatzi@psi.ch}, \email{narongrit.ritjoho@g.sut.ac.th}, \email{thomas.prokscha@psi.ch}, \email{hubertus.luetkens@psi.ch}, \email{alex.amato@psi.ch}}

\begin{abstract}
The usage of muonic X-rays to study elemental properties like nuclear radii ranges back to the seventies.
This triggered the pioneering work at the Paul Scherrer Institute (PSI), during the eighties, on the Muon Induced X-ray Emission (MIXE) technique for a non-destructive assessment of elemental compositions.
In the recent years, this method has seen a rebirth, improvement and adoption at most muon facilities around the world.
Hereby, the PSI offers unique capabilities with its high-rate continuous muon beam at the \gls{smus}.
We report here the decision making, construction and commissioning of a dedicated MIXE spectrometer at PSI, the \acrfull{mexp} setup.
Multiple campaigns highlighted the outstanding capabilities of MIXE at PSI, \eg resolving down to \SI{1}{\atp} elemental concentrations with as little as \SI{1}{\hour} data taking, measuring isotopic ratios for elements from iron to lead, and characterizing gamma rays induced by muon nuclear capture.
On-target beamspots were characterized with a dedicated charged particle tracker to be \SIlist{22.06+-0.18;14.45+-0.06}{\mm} for \SIlist{25;45}{\MeVc}, respectively.
Advanced analysis of the High Purity Germanium (HPGe) signals further allows to improve energy and timing resolutions to $\sim$\SI{1}{\keV} and \SI{20}{\ns} at \SI{1}{\MeV}, respectively.
Within the \gls{mexp} setup, an average detector has a photopeak efficiency of $\overline{\epsilon_E} = \SI{0.11}{\percent}$ and an energy resolution of $\overline{\sigma_E} = \SI{0.8}{\keV}$ at $E=\SI{1000}{\keV}$.
The overall performance of the \gls{mexp} setup at \gls{smus} allowed to start a rich user program with archaeological samples, Li-ion battery research, and collaboration with industry.
Future improvements will include a simulation based analysis and a higher degree of automation, \eg automatic scans of a series of muon momenta and automatic sample changing.

\end{abstract}

\newpage
\section{\label{sec:intro}Introduction}

The field of elemental composition analysis can draw from a rich  set of techniques.
However, there is a lack of depth-dependent analysis beyond the surface, \ie below the first $\SI{100}{\micro\meter}$.
While one can cut or destroy an object of interest and fall back to surface sensitive methods, this is clearly not an option for numerous objects, \eg precious samples such as cultural heritage artifacts or in-situ and operando studies of Li-ion batteries.
Muon Induced X-ray Emission (MIXE) however, is a non-destructive technique with access to the bulk and depth-profiling capabilities~\cite{nin2015,ha19,cl19,ni19,ge21}.

The required negative muon beams only exist at a few large-scale accelerator facilities; 1.) Paul Scherrer Institute (PSI), Switzerland; 2.) ISIS, Rutherford Appleton Laboratory (RAL), United Kingdom; 3.) TRI University Meson Facility (TRIUMF), Canada; 4.) Japan Proton Accelerator Research Complex, MUon Science Establishment (J-PARC MUSE), Japan; and 5.) MUon Science Innovative Channel (MuSIC), Japan.
Among those, a proof-of-concept study showed that the continuous high-flux Swiss Muon Source (\gls{smus}) at PSI is highly suited to host an experimental MIXE station~\cite{biswas2022}.
Therefore, the Laboratory of Muon-Spin Spectroscopy (LMU) at PSI set out to develop a dedicated experimental station.
The goal was to build a platform solution to fulfil the following conditions:
\begin{itemize}
    \item can be used in different configurations, \eg number of detectors, 
    \item a well-defined geometry for reproducibility (in following campaigns and simulations)
    \item operation as a non-stationary experiment
    \item high beam(-time) utilization
\end{itemize}

The resulting setup presented here, the \acrfull{mexp}, marks the continuation of the development of state-of-the-art equipment like the \musr{} spectrometers at PSI by LMU~\cite{lmu}.

\section{\label{sec:mixe}Muon Induced X-ray Emission (MIXE)}

The interaction of negative muons with matter provides unique elemental sensitivity.
Muons are transported by static electromagnetic optics and implanted into an object with a given momentum (typically \SIrange{20}{50}{\MeV\per{c}}).
When implanting negative muons in matter, the first deceleration processes taking place are similar to those for positive muons, \ie electrostatic interaction with the outer electrons of the target atoms.
This leads, for a large range of momentum, to a similar stopping power as the one for positive muons.
Hence negative muons
exhibit a proton-like Bragg peak stopping profile.
The implantation depth spans from as little as \SI{0.01}{\cm} for \SI{20}{\MeV} and heavy elements like lead up to \SI{1}{\cm} for \SI{50}{\MeV} and typical plastics~\cite{gr01}. 
The mean depth of resulting distributions is predominantly dependent on the object's density $\rho$, but also on its heterogeneity $\rho(\mathbf{x})$. 
The spread of the implantation profile (straggling) is additionally dependent on the muon beam momentum distribution.
Common achievable depth resolutions range from \SIrange{50}{1000}{\um}, whereby a better resolution can be achieved at a cost of reduced muon beam intensities, \ie longer acquisition times.

A typical beam spot diameter, \ie sample area illuminated, is of the order of \SI{2}{\cm}. The setup imposes little constraints on the upper limit of overall object size. Typically, the samples are mounted in air.
However, measurements in different gas atmospheres or vacuum are also possible.
Smaller samples with areas down to \si{\mm} can be measured through collimation of the beam with the downside of a lower signal rate.

At the end of the stopping process in the material, \ie when the velocity of the negative muon becomes similar to the one of the atomic electrons, the muon will interact with the Coulomb field created by the positive charge of a nearby nucleus and will be captured forming a so-called muonic atom.
The muonic atom is formed in an excited state and will rapidly ($<\SI{1}{\ns}$) emit multiple X-rays arising from the steps taken by the muon to cascade down the various energy levels~\cite{me01} until it reaches its 1$s$ ground state.
At this point, the muon either naturally decays emitting an electron and neutrinos or, in particular in high-$Z$ elements, it undergoes a nuclear capture followed by neutron and/or gamma-ray emission of the excited daughter nucleus~\cite{su87}.
The energies of the X-rays emitted during the cascade vary enough between elements, such that one can differentiate them using precision equipment.
In an approximation for muonic X-ray energies one can scale the Bohr model transition energies between principle quantum states $n$ and $m$ for electronic transitions by the ratio of muon $m_\upmu \approx 207 m_\text{e}$ and electron mass $m_\text{e}$,
\begin{equation}
E_{n\to m,\upmu} \simeq \frac{m_{\upmu}}{m_{e}}E_{n\to m,e} \simeq 207 \times E_{n\to m,e}~.
\label{eqn:eq1}
\end{equation}
The effective energy range being studied within a MIXE analysis thus spans from \SIrange{10}{10000}{\keV}.
Hence, the X-rays produced during the cascade are usually energetic enough to escape from the sample itself and can be therefore detected by detectors.
To acquire a spectrum of these photons, which are like fingerprints of elements, High Purity Germanium (HPGe) detectors are positioned around the sample~\cite{biswas2022}.
The usage of HPGe detectors permits to cover the whole energy range and provides MIXE a unique sensitivity over nearly the full periodic table. 

Another feature of MIXE is the ability to distinguish individual isotopes.
Heavy elements like Ag have line splitting due to isotope shifts on the order of several \si{\keV} and are easily detectable. 
Even for lighter elements, recent advances with precise higher order calculations of the transition energies \cite{2021_Sturniolo} open up the possibility to analyse isotopes with overlapping lines.
More detail and example is given in \cref{ssub:isotopes}.

\begin{figure}[]
    \centering
    \adjincludegraphics[width=1.0\linewidth, Clip={0\width} {0\height} {0\width} {0\height}]{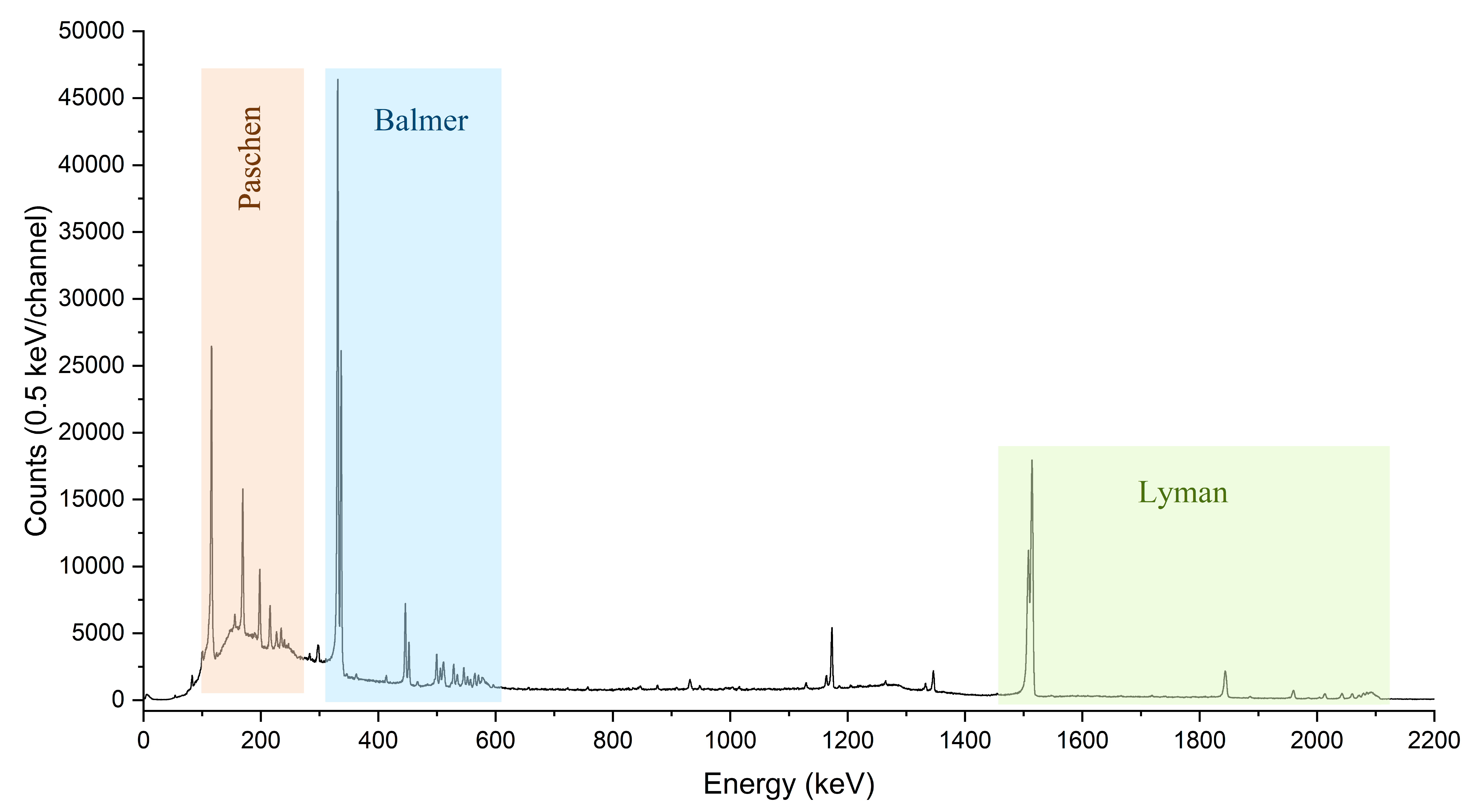}
    \caption[MIXE spectrum]{
        Example MIXE spectrum of pure Cu. The muonic Lyman, Balmer and Paschen series are clearly visible, corresponding to final muonic principal quantum number 1, 2 and 3, respectively.
    }
    \label{fig:ex_mixe}
\end{figure}

An example spectrum in Cu is shown in \cref{fig:ex_mixe}, where peaks corresponding to different transition series are clearly visible.
To acquire such spectra, the probability to measure the emitted muonic X-ray photons is a convolution of multiple factors.
Depending on its energy, a photon will experience an attenuation by the sample itself, which will depend on the elements and density $\rho(\mathbf{x})$ of the sample.
Moreover, the geometrical configuration of the sample and the detectors may lead to different path lengths for the X-rays through the sample and solid angles covered by the detectors.
Last but not least, the detection efficiency and resolution of the detector is a function of the photon energy and require proper calibration protocols.

To assess the elemental composition of an object there are generally two paths.
One has to either apply several corrections, or work with a set of reference samples such that the effects mostly cancel out.
More details are presented in \cref{sec:anaylsis}.

\section{\acrfull{smus}}

The \gls{smus} at the Paul Scherrer Institute (PSI, Switzerland) encompasses the muon beamlines and instrument complex at PSI.
It is fed by the three stage High Intensity Proton Accelerators (HIPA) complex.
The protons start with a pre-acceleration to \SI{870}{\keV} before being boosted to \SI{72}{\MeV} in a first cyclotron.
Heart of the HIPA is the final acceleration with the main ring cyclotron to \SI{590}{\MeV}.
With proton currents of up to \SI{2.4}{\milli\ampere}, it is one of the most powerful (\SI{1.4}{\mega\watt}) accelerators in the world.

\begin{figure}[]
    \centering
    \begin{subfigure}[b]{\linewidth}
        \adjincludegraphics[Clip={0.00\width} {0.00\height} {0.00\width} {0.00\height},width=1.0\linewidth]{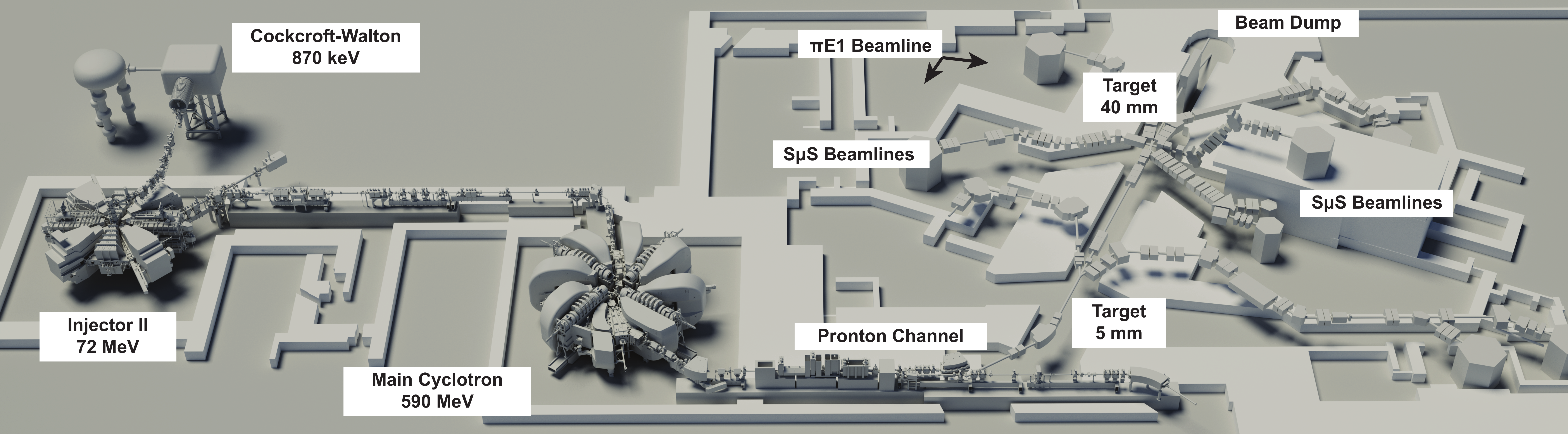}
        \caption[HIPA complex]{
            Schematic of the experimental hall and HIPA complex at PSI (Courtesy Mahir Dzambegovic, PSI).
        }
        \label{fig:hipa}
    \end{subfigure}%
    \hfill
    \begin{subfigure}[b]\linewidth
        \centering
        \adjincludegraphics[Clip={0.00\width} {0.00\height} {0.00\width} {0.00\height},width=1.0\linewidth,angle=0]{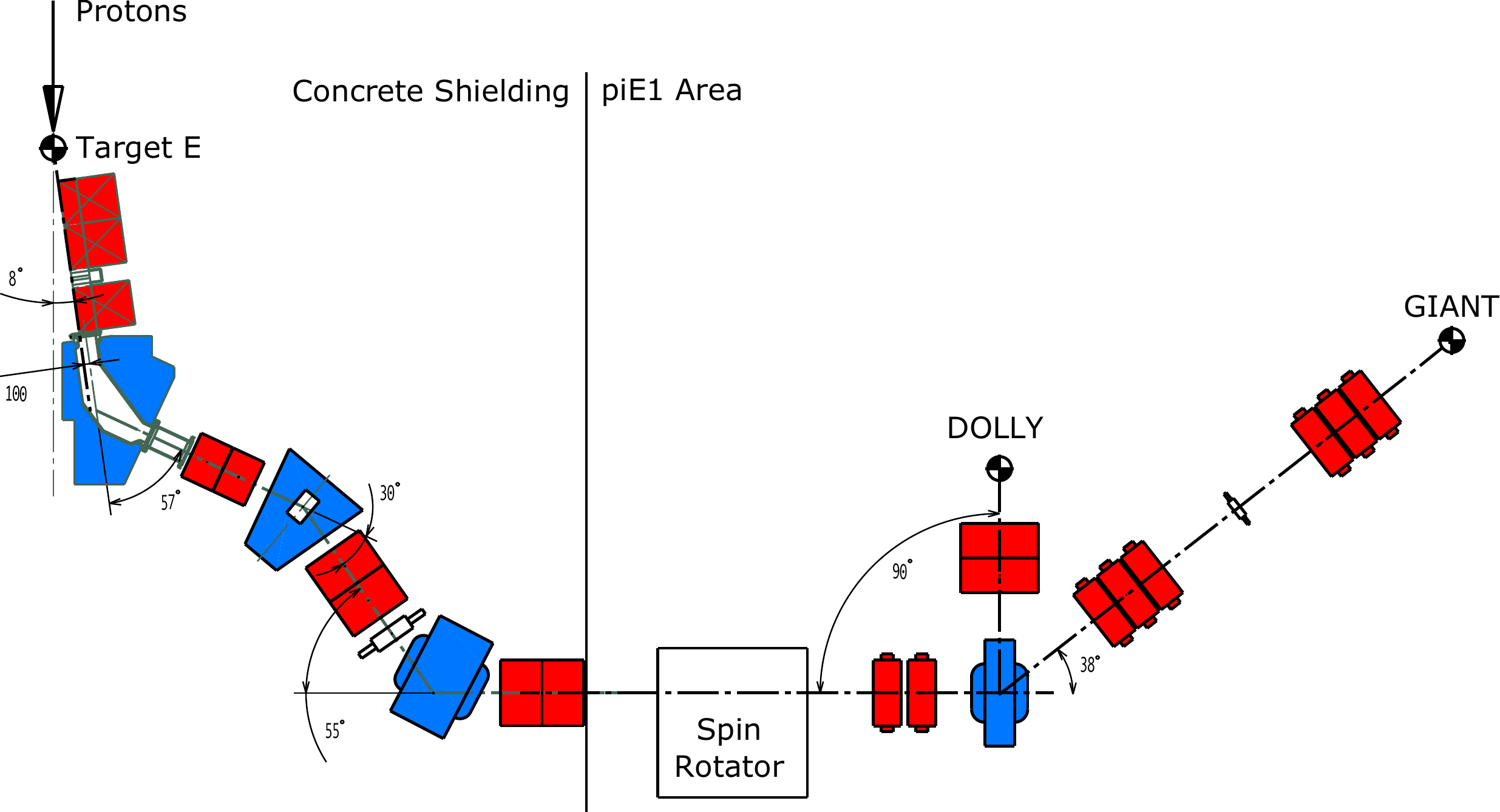}
        \caption[$\uppi$E1 beamline]{
            Layout and elements of the $\uppi$E1 beamline of the \gls{smus}. Red elements are quadrupoles, blue are dipoles, the white square marks the position of the spin rotator used as a Wien filter and the white rectangles are slit elements. Not shown is the slit element for the momentum bite selection sitting inside the second \ang{30} dipole of the early W section.
   }
        \label{fig:pie1}
    \end{subfigure}%
    \caption[\gls{smus} facilities]{
        \gls{smus} facilities at PSI used for the MIXE technique.
        }
    \label{fig:smus}
\end{figure}

The protons are utilized for secondary particle production of neutrons, pions, and muons by the subsequent decay of the pions.
Among the several beamlines capturing the produced particles, the five beamlines of the \gls{smus} ($\uppi$M3, $\uppi$E1, $\uppi$E3, $\upmu$E4, and $\upmu$E1) hosting the experimental stations of the Laboratory of Muon Spin Spectroscopy (LMU) sit at the two pion/muon production targets E and M.
A schematic of the area is shown in \cref{fig:hipa}.


\subsection{S$\mu$S $\pi$E1.2 beamline}


All MIXE campaigns at PSI so far have been conducted at the $\uppi$E1.2 beamline.
The beamline can be operated both for negative and positive decay muons and pions and has two branches through the last three port dipole magnet.
It hosts the permanently installed DOLLY \musr{} spectrometer in the {\uppi}E1.1 area at the \ang{90} port \cite{dolly}.
Using the second {\uppi}E1.2 area, at the \ang{38} branch, temporary experiments can also be installed.
The current layout of the beamline is shown in \cref{fig:pie1}.
After capturing muons from the production target, a W-section of quadrupoles and dipoles elements allows the operation of an overall achromatic mode with a momentum dispersion within the second dipole.
This second \ang{30} dipole has an integrated horizontal slit system that is used to block parts of the muon beam, effectively reducing the momentum bite $\Delta p/p$ by sacrificing rate.
After the W-section a spin rotator with up to $\pm\SI{250}{\kV}$ serves not only for spin rotation for \musr~experiments but also as a Wien filter, reducing the electron or positron contamination.

\begin{figure}[]
    \centering
    \adjincludegraphics[width=0.65\linewidth, Clip={0\width} {0\height} {0\width} {0\height}]{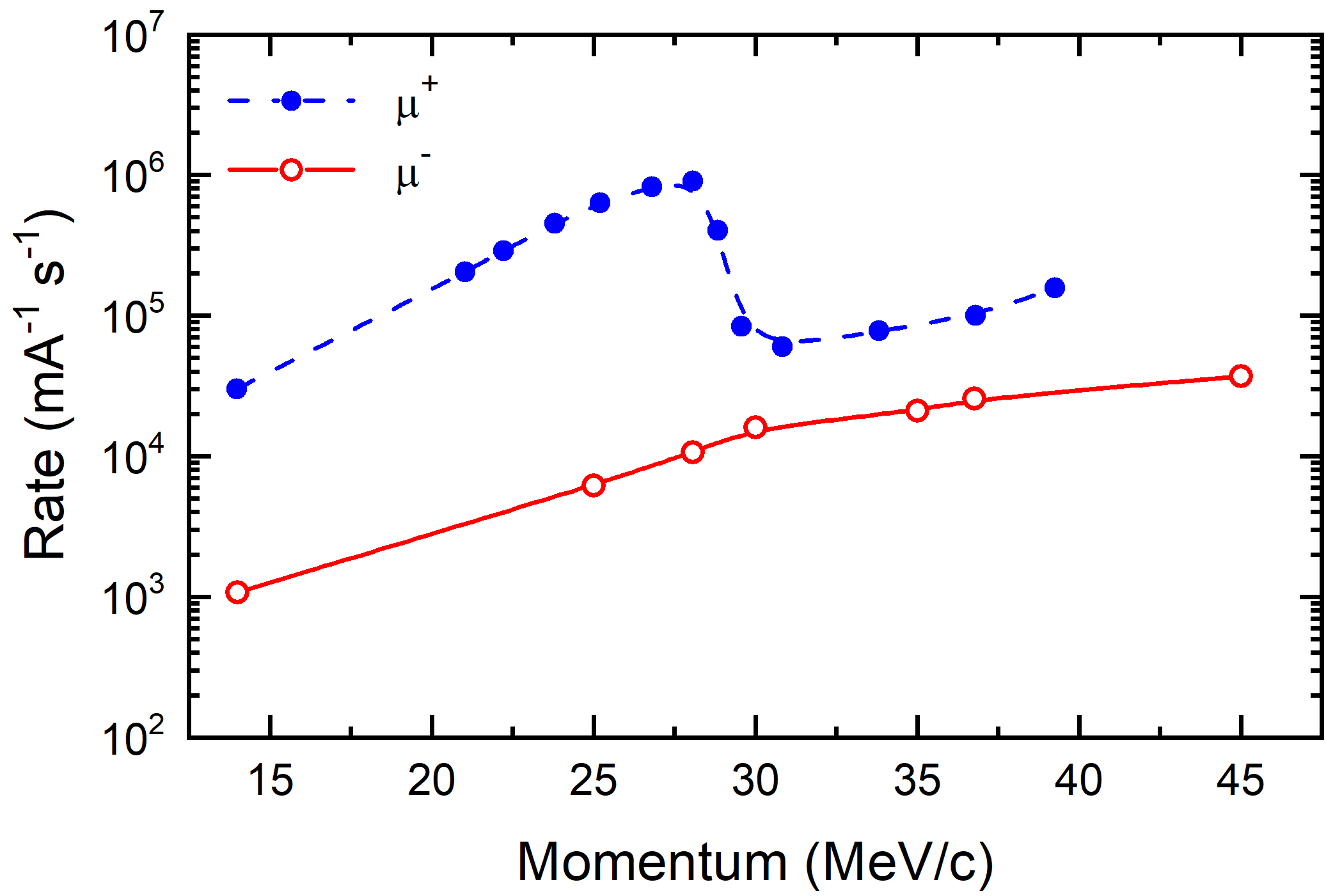}
    \caption[$\upmu^-$ rates]{  
        Muon rates at the $\uppi$E1 beamline for different momenta. The momentum bite is 2\% here.
    }
    \label{fig:rates}
\end{figure}

Momenta for negative muons with acceptable rates are between \SIrange[list-separator = { and }, per-mode = symbol]{20}{50}{\MeVc} (see Fig.~\ref{fig:rates}).
This corresponds to a penetration of \SIrange[per-mode = symbol]{.04}{1}{\gram\per\square\cm}.

\subsection{Beam spot}

\begin{figure}[]
    \centering
    \adjincludegraphics[width=0.8\linewidth, Clip={0\width} {0\height} {0\width} {0\height}]{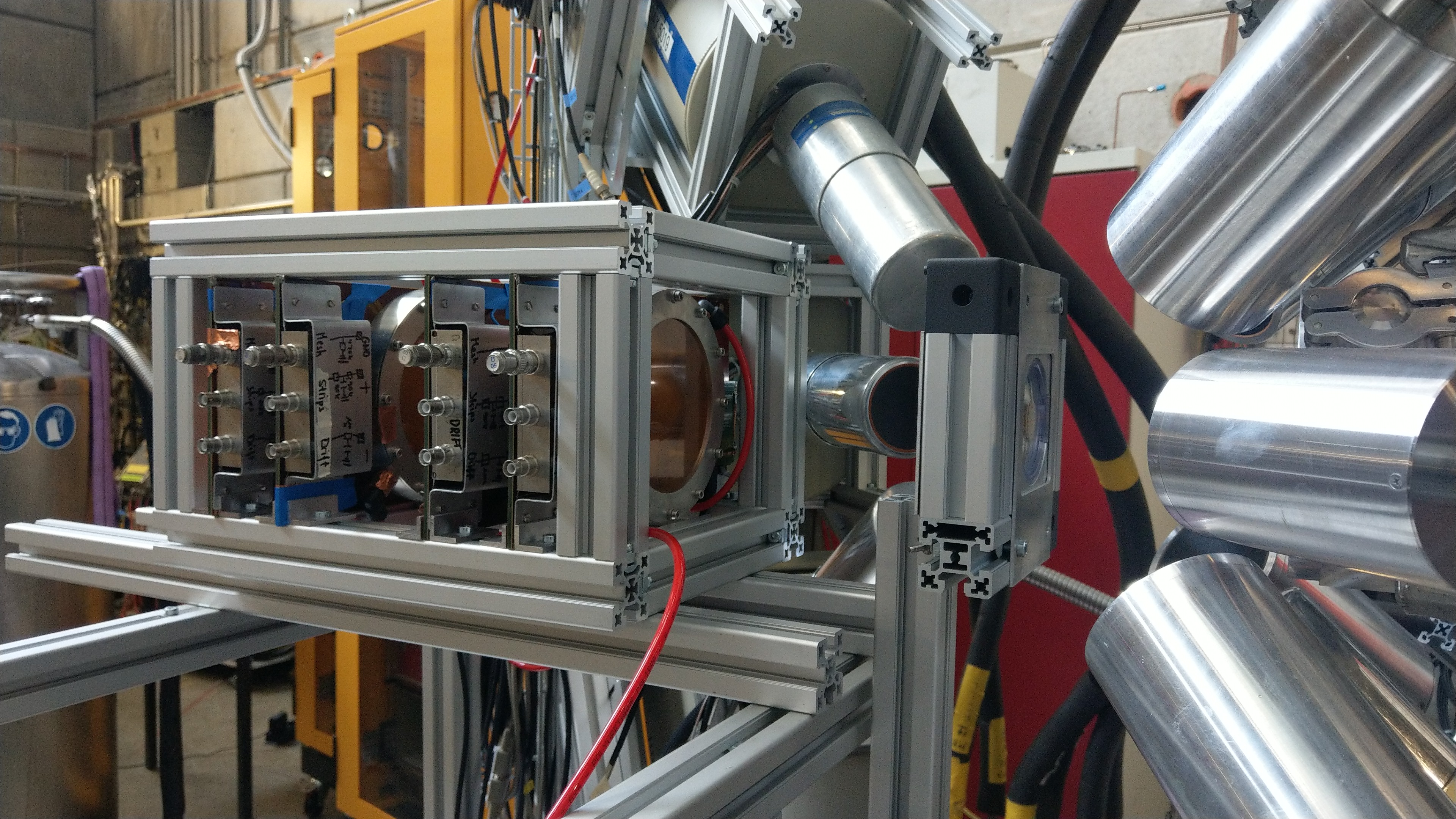}
    \caption[Pictures of micromegas setup]{
        Picture of the setup of micromegas telescope downstream of the sample.
    }
    \label{fig:micromegas}
\end{figure}

The Micromegas (MICRO MEsh GAseous Structure) is a type of micropattern gaseous detector, invented by I.~Giomataris and G.~Charpak in \num{1995}  \cite{1996_Giomataris}.
To study the size of the muon beam spot at the sample position, four Micromegas detectors (MMDs) arranged in a telescope were set up (see Fig.~\ref{fig:micromegas}).
These MMDs have an active area of \num{80}x\SI{80}{\milli\meter\squared} and were originally used by the NA64 experiment at CERN \cite{2018_Banerjee}.
The MMDs were continuously flushed with a gas mixture of Argon/CO2 (\SI{82}{\percent}/\SI{18}{\percent}). 

\begin{figure}[]
    \centering
    \adjincludegraphics[width=0.9\linewidth, Clip={0\width} {0\height} {0\width} {0\height}]{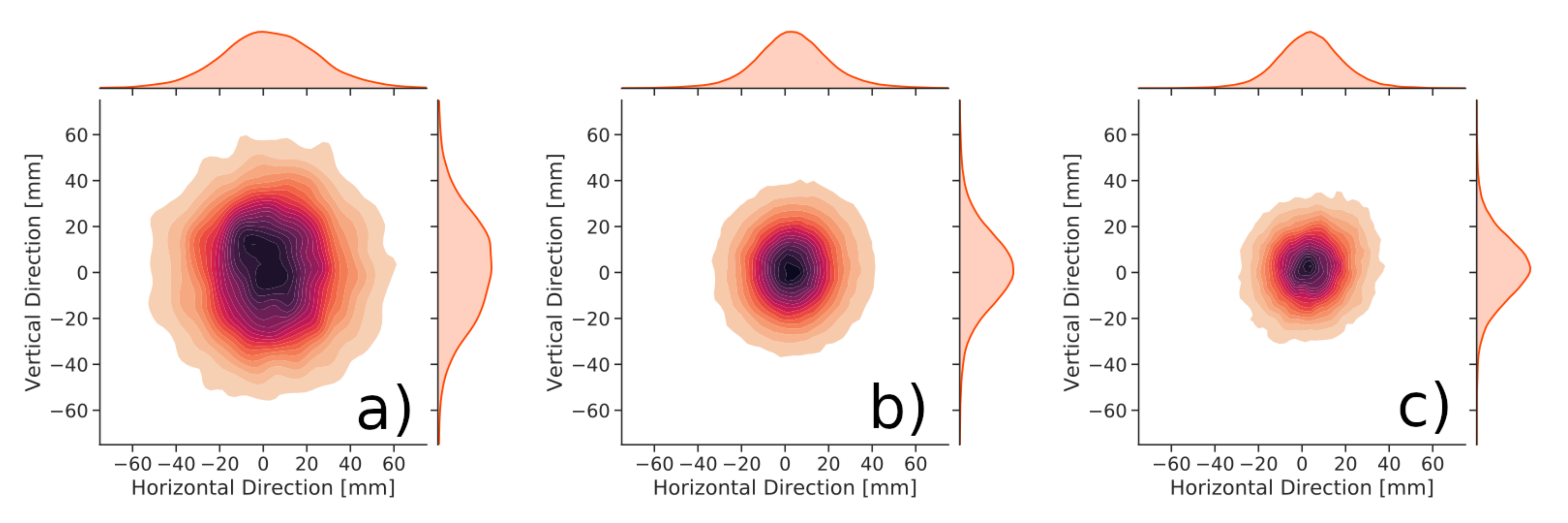}
    \caption[Beam spot]{
        Beam spots on target during the MIXE campaign in 2022 May in $\uppi$E1 for different momenta: (a)~\SI[per-mode = symbol]{25}{\MeVc}, (b)~\SI[per-mode = symbol]{35}{\MeVc} and (c)~\SI[per-mode = symbol]{45}{\MeVc} .
    }
    \label{fig:beamspot}
\end{figure}

In order to reconstruct the beam spot, the muon beam was stopped at the sample position in a beam dump.
Once the muons decayed, their decay-electrons then traversed through the telescope and ionized the gas mixture.
The freed electrons were guided by an electrical field of the order of \SI{500}{\volt\per\centi\meter} through the micro-mesh into a second region, where a strong electrical field was applied (\SI{50}{\kilo\volt\per\centi\meter}).
The free electrons were accelerated and gained enough energy to further ionize the gas, kicking off an avalanche process.
At the end of this region the local avalanche reached two differently oriented planes (X and Y) with each \num{320} readout strips, where the amplified signal could be read out, allowing for a hit reconstruction in 2D with a spatial resolution of the order of \SI{100}{\micro\meter}.
When at least three out of four detectors of the telescope had a hit recorded, a track could be fitted and extrapolated onto the sample position.
The beam spot was reconstructed for different momenta, shown in Fig.~\ref{fig:beamspot}.
By fitting a gaussian distribution in both the horizontal and vertical axis of the beam spot, its standard deviation was taken as figure of merit for the beam spot size.
The results are summarized in \cref{tab:summary_beamspot}.

\begin{table}[!htbp]
    \sisetup{
    }
  \centering
  \caption[Summary of the beam spot sizes at different momenta]{Summary of the beam spot sizes in horizontal ($\sigma_x$) and vertical ($\sigma_y$) direction at different momenta.}
    \begin{tabular}{ccc}
    \hline \noalign{\vspace{-1pt}} \hline
    \multicolumn{1}{c}{Momentum}&
    \multicolumn{1}{c}{$\sigma_x$} & 
    \multicolumn{1}{c}{$\sigma_y$}\cr
    
    \multicolumn{1}{c}{(\si{\mega\electronvolt\per\c)}}&
    \multicolumn{1}{c}{(\si{\milli\meter})} & \multicolumn{1}{c}{(\si{\milli\meter})}\\
    
        \noalign{\smallskip}
         \hline
        \noalign{\smallskip}
\num{25} & \num{22.06+-0.18} & \num{23.54+-0.18} \cr       
\noalign{\smallskip}
\num{33} & \num{17.52+-0.03} & \num{18.07+-0.03}  \cr
        \noalign{\smallskip}
\num{35} & \num{16.55+-0.03} & \num{17.24+-0.03}  \cr
        \noalign{\smallskip}
\num{45} & \num{14.45+-0.06} & \num{14.34+-0.06}  \cr
        \noalign{\smallskip}
           \hline \noalign{\vspace{-1pt}} \hline
    \end{tabular}%
  \label{tab:summary_beamspot}%
\end{table}%

\section{The \acrfull{mexp} setup at PSI}

\begin{figure}[]
    \centering
    \adjincludegraphics[width=0.9\linewidth, Clip={0\width} {0\height} {0\width} {0\height}]{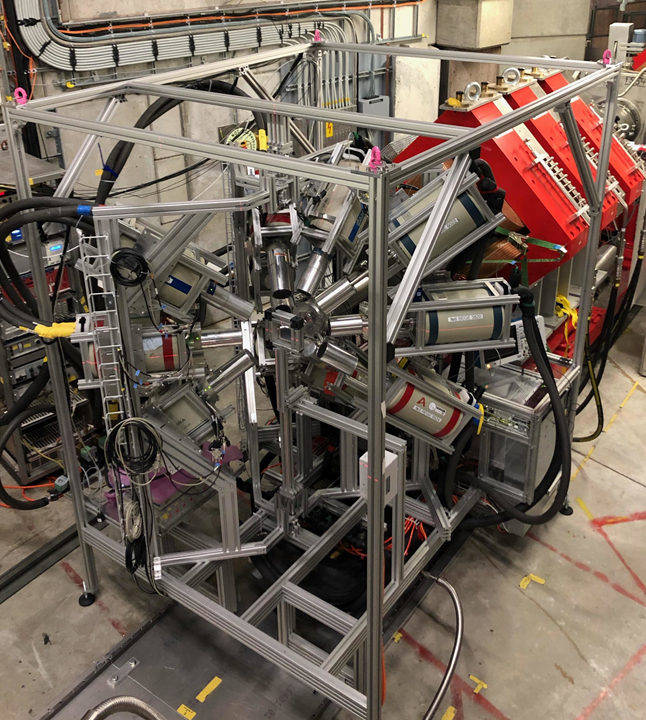}
    \caption[Picture of \gls{mexp} setup]{
        Picture of the \gls{mexp} setup in 2021.
    }
    \label{fig:dora_pic}
\end{figure}

The \acrfull{mexp} setup was built under the constraints mentioned in the introduction \cref{sec:intro}.
Not being a permanently installed experimental setup at PSI, but only used during beamtimes of a week length, quick setup times for high beamtime utilization was required.
Practically, it meant to develop a platform where electronics, cabling, and options for alignment of the detectors and the overall setup relative to the beam could be prepared offline prior to the actual beamtimes.
The requirement of reproducibility comes in twofold.
First, a preparation including cooling of the detectors and calibration of the setup outside the experimental area, transportation, and recommissioning requires to separate the fragile HPGe detectors from the setup for transportation.
Hence, a quick and straight forward re-installation of the detectors at identical positions is compulsory.
Second, to be able to correct for the effects on X-ray spectrum acquisition discussed in \cref{sec:mixe}, precise knowledge about the positions of the detectors is required.
Last but not least, an optimum use of the beamtimes meant to reduce the duration of costly interruptions, \eg sample changes and detector refill.

\subsection{Experimental setup}

The primary considerations for a MIXE experimental setup are the arrangements of the sample to the beam(port), and the detectors to the sample.
As most samples are measured in air, a minimal distance between beam window and sample to reduce the scattering of muons in air is desirable.
The placing of the HPGes is more complex.
A closer position increases the solid angle and thus the detection efficiency.
On the other hand, the de-excitation of the muonic atom emits multiple photons.
Therefore, the solid angle should still be small enough to maximize the chance to only have a single X-ray interacting per detector at a given time.
Depending on the muons source, pulsed or continuous beams, the co-existence of multiple muons within the acquisition window of a germanium detector ($\Delta t\approx\SI{10}{\us}$) needs to be taken into account, \ie a high instantaneous rate at a pulsed source requires a higher detector granularity/smaller solid angles.
As an approximation this can be estimated with the binomial distribution and Poisson statistics, a more detailed optimization requires detailed know-how and simulations.
For example, one can consider the following estimations.
Assume that there is only one muon $n_\upmu=1$ in a given time frame interacting with the sample. The formed muonic atom emits $n_\upgamma=10$ photons during the de-excitation.
The approximated solid angle $\Omega$ for a cylindrical detector of radius $R$ with distance $d$ to the sample in \si{\percent} can be expressed as a disc projected on and normalized by the sphere of isotropic emission
\begin{align}
    \Omega      &= 2\pi \left( 1 - \frac{d}{\sqrt[]{R^2+d^2}}\right) \times (4\pi)^{-1}~.\\
    \end{align}
Taking $\Omega=\SI{10}{\percent}$ and further assuming that the interaction probability (creation of a signal via Compton, photo effect and/or pair-production) for a photon with the detector is $\epsilon=\SI{100}{\percent}$, the chance that $i$ photons interact with a detector, $P_i$, would be
\begin{equation}
    P_i = \binom{n_\upgamma}{i}\,(\epsilon \Omega)^i\,(1-\epsilon \Omega)^{n_\upgamma-i}~.
\end{equation}
Thus, for this example the single photon interaction probability would be $P_1=\SI{39}{\percent}$.
However, this approximation ignores that the multiplicity of photons is element/sample dependent and that the interaction efficiency of the X-ray with the detector has an energy dependence (to some extent also convoluted with the solid angle).
Nevertheless, it makes sense to use such approximation to find an upper limit for the solid angle.
For the case of the \gls{mexp} setup we used the worst case scenario of $n_\gamma = 15$.
This is motivated because the typical principle muon quantum number after capture is 14,  corresponding to the square-root of the muon-electron mass ratio.
Moreover, we assume that an additional photon arises from the potential emission of nuclear gamma after the capture of the muon by the nucleus and $\epsilon = \SI{100}{\percent}$.
The detectors in the setup have radii $R\leq\SI{5}{\cm}$.
Not optimized parts of the beamline window constrain the detector sample distance to $d=\SI{15}{\cm}$.
The approximated single photon probability is 
\begin{equation}
    P_1 = \SI{27}{\percent} < \text{max}(P_1) \approx {P_1}\Bigr\rvert_{d=\SI{8.7}{\cm}} = \SI{38}{\percent}~.    
\end{equation}

When including muon rates $\Phi$ into account for pileup, the probability to measure a single photon now requires that in the \SI{10}{\us} before and after no photon interacted with the detector, denoted as the probabilities $P_{0,\text{b}}$ and $P_{0,\text{a}}$ respectively.

\begin{align}
    P' =& P_{0,\text{b}} \; P_1 \; P_{0,\text{a}}~,\\
       =& P_1 \; {P_{0,a}}^2~,\\
\end{align}
Using the relation that no interaction and any ($i>0$) interaction follow
\begin{equation}
    P_{0,a} = 1 - \sum_{i>0} P_{i,a}
\end{equation}
one can sum over the Poisson probability mass function of possible muon pile-up weighted with probability of photon interaction
\begin{align}
       P'=& P_1 \; \left[1-\sum_{n_\upmu=1}^{\infty} (1-{P_0}^{n_\upmu}) \; \text{Poiss}({n_\upmu},\Phi\Delta t)\right]^{\;2}~.\\
\end{align}
Expanding the previous worst case with a rate of $\Phi=\SI{50}{\kHz}$ one gets
\begin{equation}
       P' = \SI{27}{\percent} \; (\SI{85}{\percent})^2 = \SI{20}{\percent}~.
\end{equation}

If the measurement of low energetic X-rays is of interest, also the angle of the detector axis to the sample surface becomes increaslingly relevant.
Taking the more extreme example of measuring Li in a battery pouch cell.
For the sake of the argument other battery materials which will only make matters worse are ignored.
A typical pouch out of aluminum has a wall thickness of $\ell=\SI{100}{\um}$ with the Al density $\rho=\SI{2.7}{\gram\per\cubic\cm}$.
The prominent X-ray line of Li is $E_{\upmu,2\text{p} \to 1 \text{s}}=\SI{18.7}{\keV}$.
The mass attenuation coefficient of Al $\mu/\rho$ at this energy is \SI{4.6146}{\cm\squared\per\gram} \cite{xray}.
The total attenuation $\Gamma$ for a path at an angle $\phi$ relative to the surface towards a detector can thus be calculated as
\begin{equation}
    \Gamma(\phi) = 1 - \exp\left( - \mu/\rho \frac{ \rho\, \ell }{\cos(\phi)}\right)~.
    \label{eq:attenuation}
\end{equation}
A selection of values is shown in \cref{tab:att}.
It becomes eminent that small angles are of clear advantage and detectors at different positions require proper corrections in order to compare them.
Another feature to consider is the finite size of the detector.
A HPGe of diameter \SI{80}{\mm} at distance \SI{15}{\cm} from a sample covers an angle of $\ang{15}$
relative to its central axis.
This implies a range of attenuation for a single detector.
An average over the full solid angle of a detector needs to be made for an accurate correction.
At this point, such increased complexity of corrections shows that one has to rely on a full scale simulation with a toolbox like Geant4 \cite{geant4} to optimize a setup and/or correct the acquired data.
Lastly, care should be taken that the HPGes are not exposed to large background from beam halos, misguided or secondary particles and their bremsstrahlung.

\begin{table}
    \centering
    \caption[X-ray attenuation]{X-ray attenuation for Li at $E_{\upmu,2\text{p}\to1\text{s}}=\SI{18.7}{\keV}$ through a \SI{100}{\um} Al foil for paths to a detector with radius $r=\SI{5}{\cm}$ at $d=\SI{15}{\cm}$ distance and various angles $\phi$ relative to the foil surface.}
    \begin{tabular}{llllll}
        \toprule
        $\phi$ / \si{\degree}       & 0    & 30   & 45   & 60   & 75    \\ 
        \midrule
        $\Gamma(\phi)$ / \si{\percent} & 11.7 & 13.4 & 16.2 & 22.1 & 38.2 \\
        \bottomrule
    \end{tabular}
    \label{tab:att}
\end{table}

\begin{figure}[]
    \centering
    \begin{subfigure}[b]{0.74\linewidth}
         \adjincludegraphics[width=\textwidth, Clip={0.0\width} {0\height} {0.0\width} {0\height}]{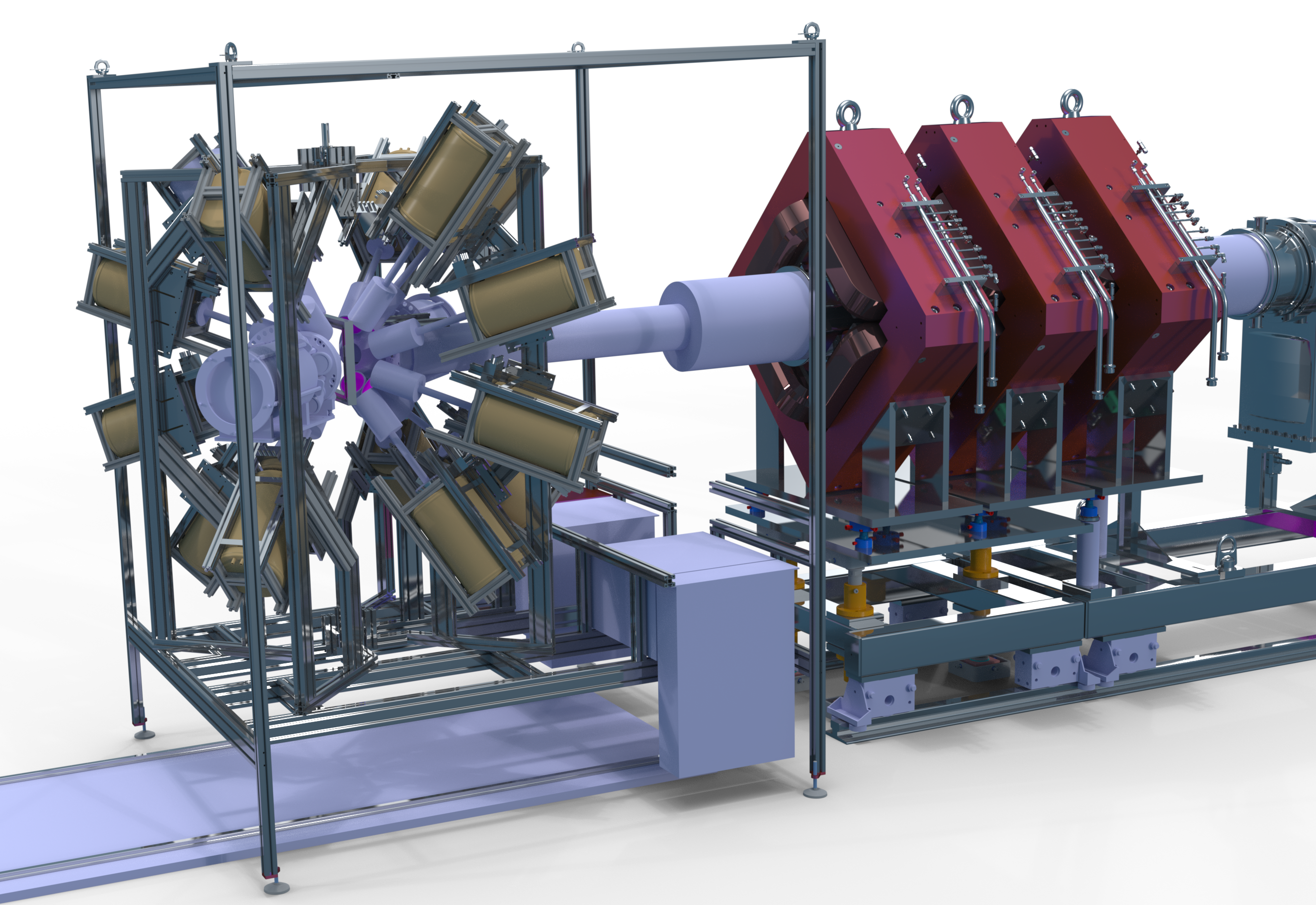}
        \caption[\gls{mexp} elements]{
            CAD model of the \gls{mexp} frame, HPGe frames, arms, sample system.
        }
    \label{fig:giant_cad}
    \end{subfigure}
    \hfill
    \begin{subfigure}[b]{0.25\linewidth}
         \adjincludegraphics[width=\textwidth, Clip={0\width} {0\height} {0\width} {0\height}]{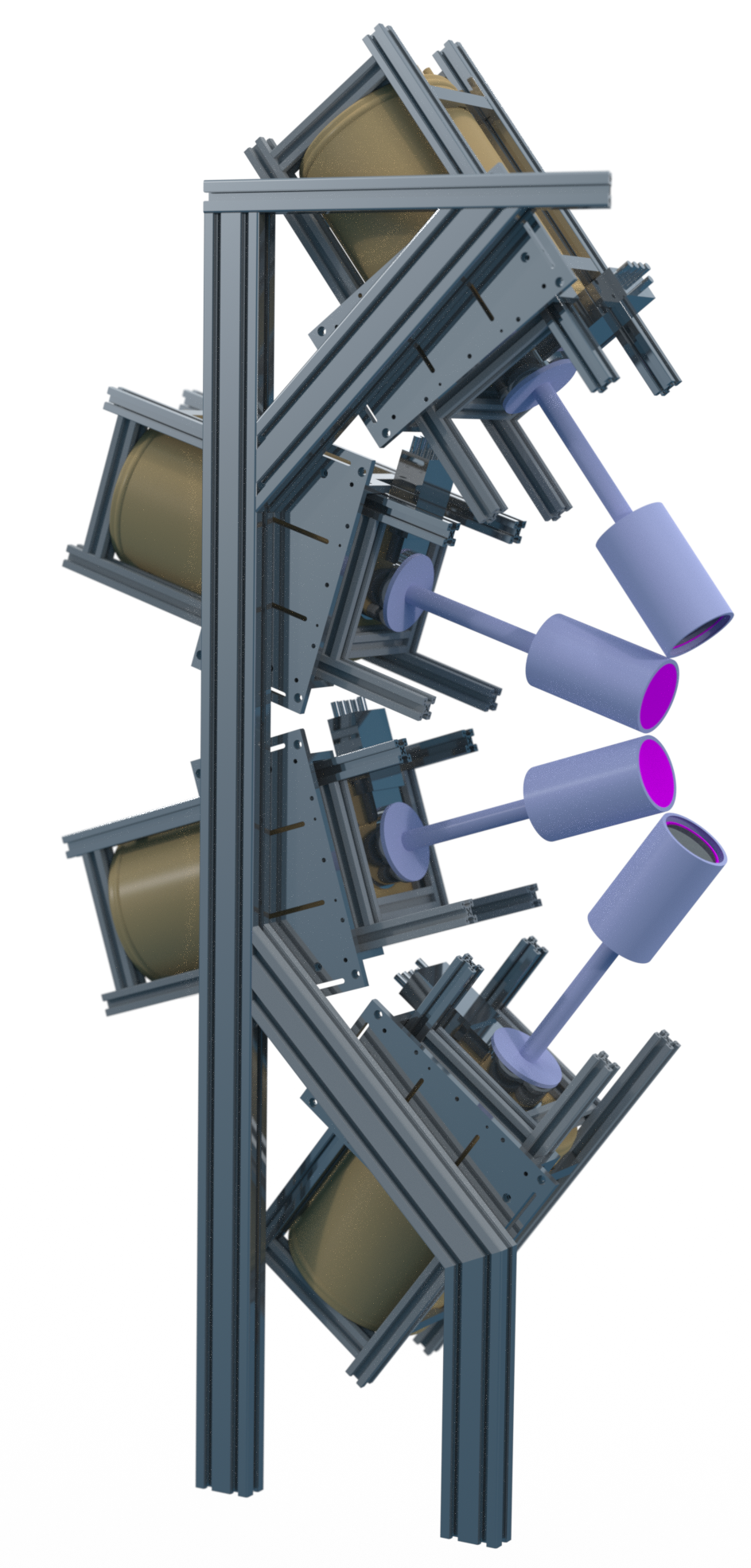}
        \caption[\gls{mexp} arm]{
            Model of a detector arm.
        }
    \label{fig:arm_cad}
    \end{subfigure}%
    \caption[\gls{mexp} CAD models]{
        CAD models of the different elements. \gls{mexp} frame, HPGe frames, arms, and sample system.
        }
    \label{fig:giant_elements}
\end{figure}

The considerations above show that for a proper optimization of a MIXE setup a set of rigorous simulations is required. 
Limited development time led to the construction of a versatile instrument which fulfils the stated requirements and allows a later optimization.
A picture of the setup is shown in \cref{fig:dora_pic}.
Most of the mechanics was done with aluminum construction profiles. 
The maximum capacity of the system is to host 22 HPGes.
The first campaign in 2021 only used 6 detectors, with later increments to 11 and 14 HPGes for the latest 2022 measurement period.
Performance and detection efficiencies are presented in \cref{sec:anaylsis}.
The used HPGe detectors are from Mirion Technologies with {7\;l} liquid nitrogren (\LN{}) \emph{BigMac} cryostats.
To have proper detection efficiency and energy resolution over the full energy range both planar (low-energy) and co-axial (high-energy) detectors are used.
Each detector is held in a frame that allows easy (dis-)mounting and distance adjustment.
The angular alignment is solved by combining multiple detectors on vertical arms.
An arm can be freely positioned at a wanted angle in the horizontal plane $\theta$ and carry up to four detectors.
The angle between two detectors in the vertical plane of an arm hast to be $\Delta\phi\geq\ang{35}$.
Maximum angles are $\phi=\pm\ang{55}$ with $\phi=0$ being horizontal.
The arms are fixed on a central axis to create a common focal point of all detectors.
Models of the individual elements are presented in \cref{fig:giant_elements}.
Such ensemble is mounted to a larger platform which also hosts the electronics and \LN{} distribution.
The platform runs on rails parallel to the beam axis to easily (dis-)engage the beamport with the whole setup.
On the start of a beamtime, the reproducibility and integration of electronics allows to push the setup time (mechanical, electrical and cryogenic setup and commissioning) down to $\approx\SI{4}{\hour}$.

To reduce the background of uncorrelated signals in the germanium detectors a typical scheme for continuous beams is to use a muon tagging.
All MIXE campaigns at PSI so far were making use of the plastic scintillator setup developed originally for the muX experiment~\cite{kn20}.
It is composed of a $\SI{200}{\um}\times\SI{25}{\mm}\times\SI{25}{\mm}$ scintillating plastic polyvinyltoluene (BC-400), called M-counter, masked with a $\SI{8}{\mm}\times\SI{10}{\cm}\times\SI{10}{\cm}$ BC-400 veto counter with a \SI{18}{\mm} central hole.
The scintillators are paired with Silicon Photomultipliers (SiPMs) and in-house developed pre-amplifiers.

With samples of various dimensions and compositions each sample requires individual alignment.
To maximize the beam utilization the time-intensive sample mounting and alignment is not performed on the instrument itself.
A copy of the sample mount and a laser mimicking the focal point of the detectors and muon beam allows to prepare the sample without interrupting the beam.
On sample change the prepared adapter only needs to be transferred to the setup in the experimental area resulting in a down-time
of less than \SI{5}{min}.

Common materials hosting the often fragile samples (\eg archaeological objects) are low density polyethylene (PE) foam, Al or Cu wires.
If a sample has (partial) transparency for muons a controlled beam dump of \SI{1}{\cm} PE or \SI{5}{\mm} Al to stop \SI{45}{\MeVc} muons can be used.
Small objects and especially for measurements of lateral sub-regions of a larger object the mounts can be combined with a collimator out of PE or Al.

The efficiency and energy calibration of the setup was done with a large set of radioactive sources in combination with measurements of pure samples.
Details are presented in \cref{sub:efficiency}.

\begin{figure}[]
    \centering
    \adjincludegraphics[width=0.9\linewidth, Clip={0\width} {0\height} {0\width} {0\height}]{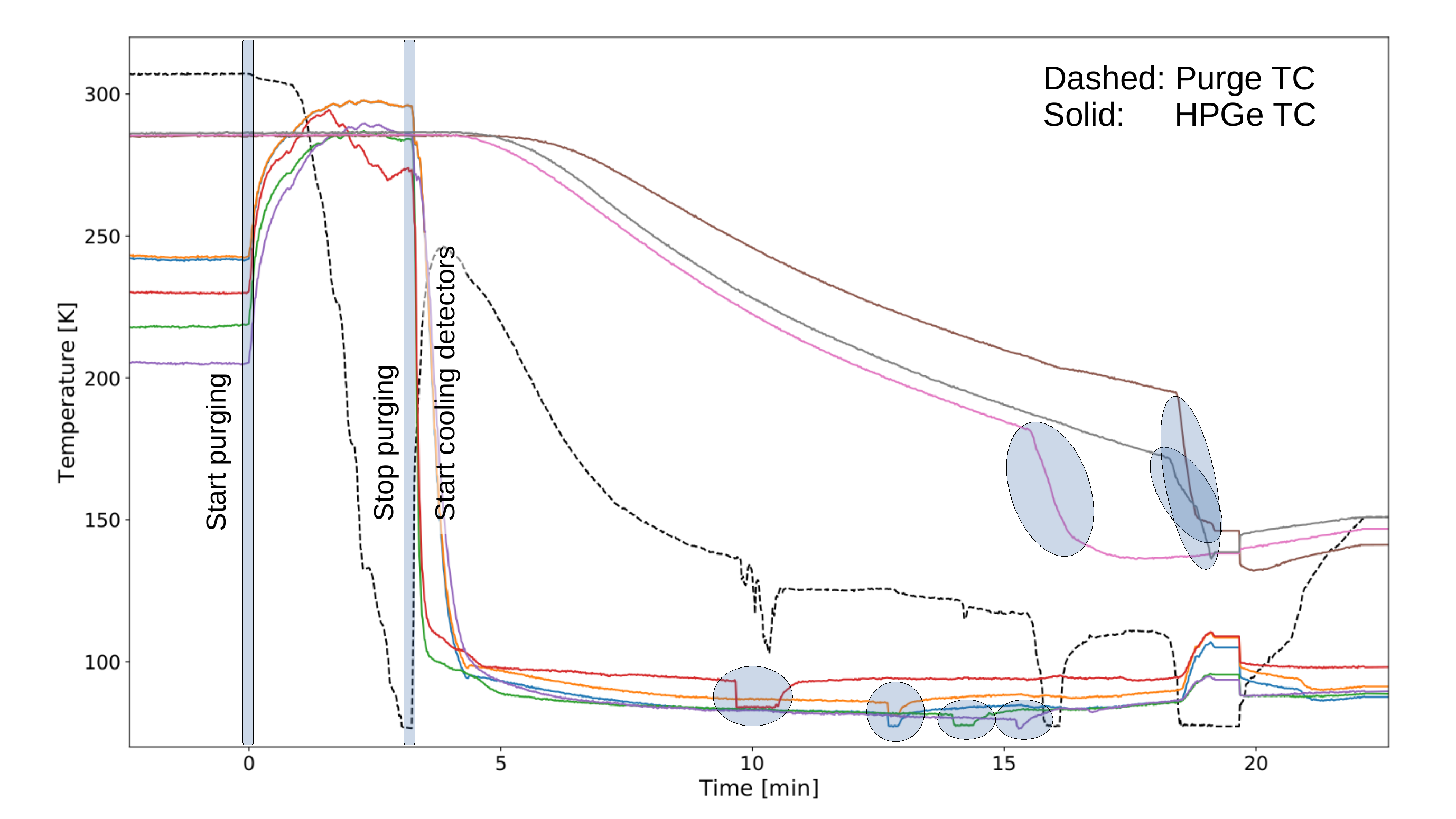}
    \caption[LN$_2$ system]{
        Picture of liquid nitrogen filling system and thermocouple readings of filling cycle. The dashed line shows the thermocouple of the purging system, whereas the solid lines are for the different HPGe detectors. The circles are marking the time when a specific detector is completely filled.
    }
    \label{fig:ln2}
\end{figure}

Another potential beamtime interruption is the filling of the \LN{} dewars of the HPGe detectors.
Remedy comes by remote control of the refilling process.
The cryostats are connected through a network of relay switched solenoid valves to a 200\;l storage dewar at \SI{1.5}{\bar}.
To monitor the filling process K-type thermocouples installed in the \LN{} return lines serve as a cost effective reading to distinguish N$_2$ from \LN{}.
A timeseries of the filling process and the temperature calibrated thermocouple values is shown in \cref{fig:ln2}.

\subsection{Data acquisition}

All detectors, HPGe and plastic scintillators, are connected to 16-channel SIS3316 \SI{14}{\bit} \SI{250}{\mega\Samples} VME digitizers from Struck Innovative Systems (SIS).
The communication with the DAQ computer is done via glass fiber between a VME communication module SIS3104 and a PCI card SIS1100.
All software front- and backends are based on the Maximum Integrated Data Acquisition System (MIDAS)~\cite{midas}. 
Signals from the detectors are passed through a trapezoidal filter on the FPGA chip of the SIS3316 module with parameters optimized for each channel to provide ADC values related to the energy of the signal.
In parallel, the rising edge of the HPGe waveforms is digitized for $300-400\times\SI{4}{\ns}$ samples to allow later timing, pile-up and energy corrections.
Such corrections and coincidence handling between plastic scintillators and HPGes is performed offline on another analysis server.

\section{\label{sec:anaylsis} Data Analysis}

\subsection{\label{sub:efficiency}Energy and efficiency calibrations}
\begin{figure}[ht]
    \centering
    \adjincludegraphics[width=1.0\linewidth, Clip={0\width} {0\height} {0\width} {0\height}]{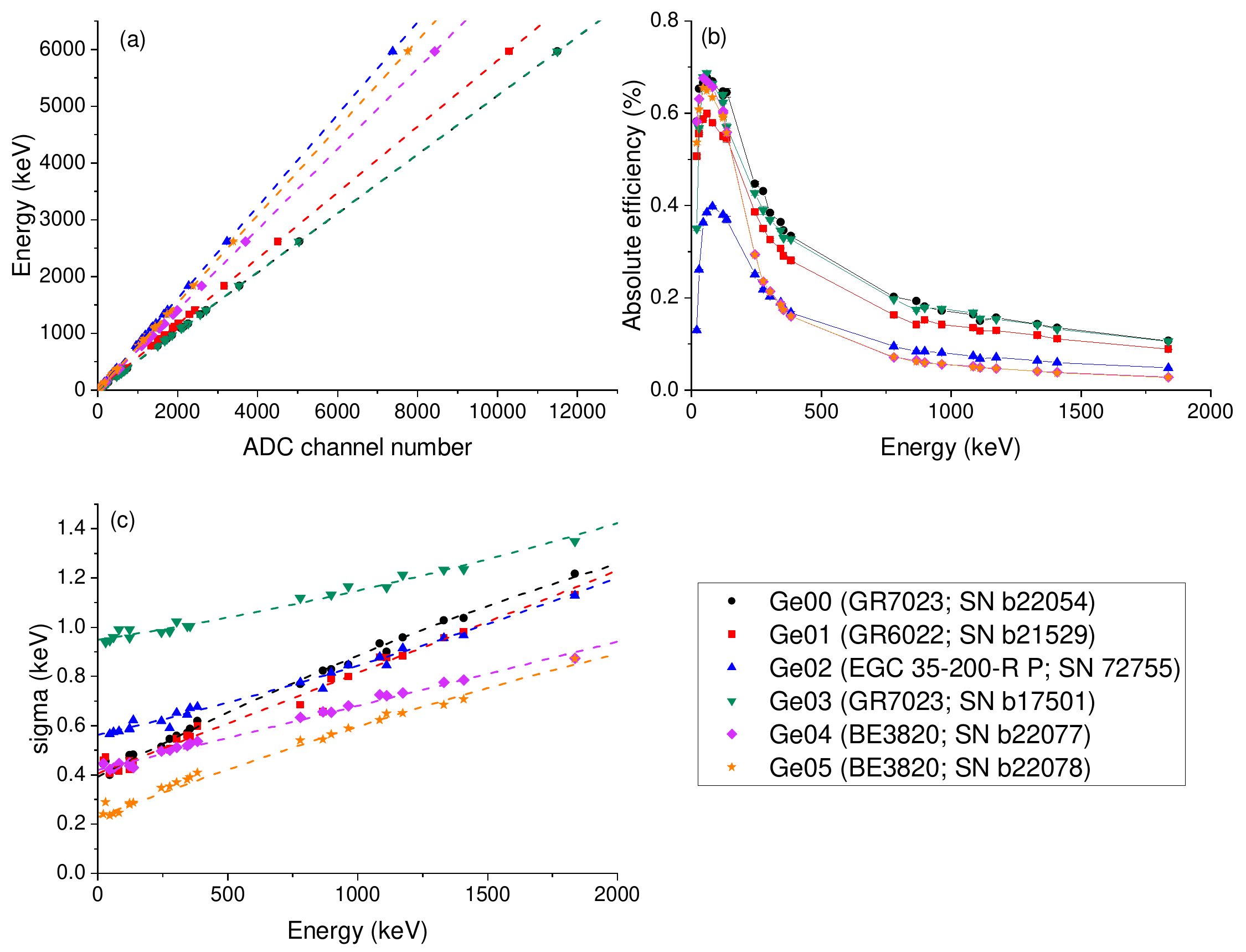}
    \caption[Calibration]{Plots of (a) energy calibration, (b) efficiency calibration and (c) standard deviation of the gamma peaks for the different detectors used in the October 2021 MIXE campaign.
    }
    \label{fig:calibration}
\end{figure}

The energy and efficiency calibrations of the HPGe detectors are performed using standard radioactive gamma-emitting sources with known activities and reference dates.
As an example, the calibrations performed during the first MIXE campaign in 2021, when there were six HPGe detectors placed in the \gls{mexp} setup, is shown in Fig.~\ref{fig:calibration}.
The distance between the focal point and the surface of the detectors was $\sim$145~mm.
The standard radioactive sources (with 3\% error on the activities from the manufacturer) of $^{88}$Y, $^{152}$Eu, $^{241}$Am, $^{210}$Pb, $^{60}$Co, $^{133}$Ba, $^{57}$Co, and $^{109}$Cd were used for calibration purposes.  
With these sources, the calibrations could be performed only up to $\sim$1.8 MeV. 
Hence the 2614.511(10)~keV of $^{208}$Tl (produced during the decay of $^{228}$Th), present in the natural background run, was also taken into account.
In addition, the $K_{\alpha 1}$ muonic X-ray of 208Pb (present in natural-Pb) at 5963.77(45) keV (determined precisely in 1970's~\cite{1969_Anderson}) was used to extend the calibration points up to $\sim$6.0~MeV.
Figure~\ref{fig:calibration}(a) shows the energy calibration data points along with the fits (using a quadratic polynomial function) for the six detectors, the details of which are shown in the figure legend.
The data points for absolute efficiency and the sigma of the peaks (fit with a Gaussian function) till $\sim$1.8 MeV are shown in Figs.~\ref{fig:calibration}(b) and (c), respectively.
The data points in Fig.~\ref{fig:calibration}(c) are also fit with a quadratic polynomial function.
From Fig.~\ref{fig:calibration}(b), it is evident that the "GR" detectors have higher efficiency than the "BE" detectors at higher energies.
However, the "BE" detectors have much better resolution than the "GR" detectors at all energies, as evident from the plots in Fig.~\ref{fig:calibration}(c).

\subsection{\label{ssub:corrections} Pre-analysis corrections of in-beam data}
In the offline analysis, before analyzing the in-beam spectra from actual samples, two different corrections are employed, by using the digitized preamplifier pulses of each HPGe detector: (i) baseline corrections and (ii) timing corrections.
The details on these corrections can be found in Ref.~\cite{alex_thesis}.


\subsubsection{\label{ssubsub:baseline_corrections} Baseline corrections:}
\begin{figure}[ht]
    \centering
    \adjincludegraphics[width=1.0\linewidth, Clip={0\width} {0\height} {0\width} {0\height}]{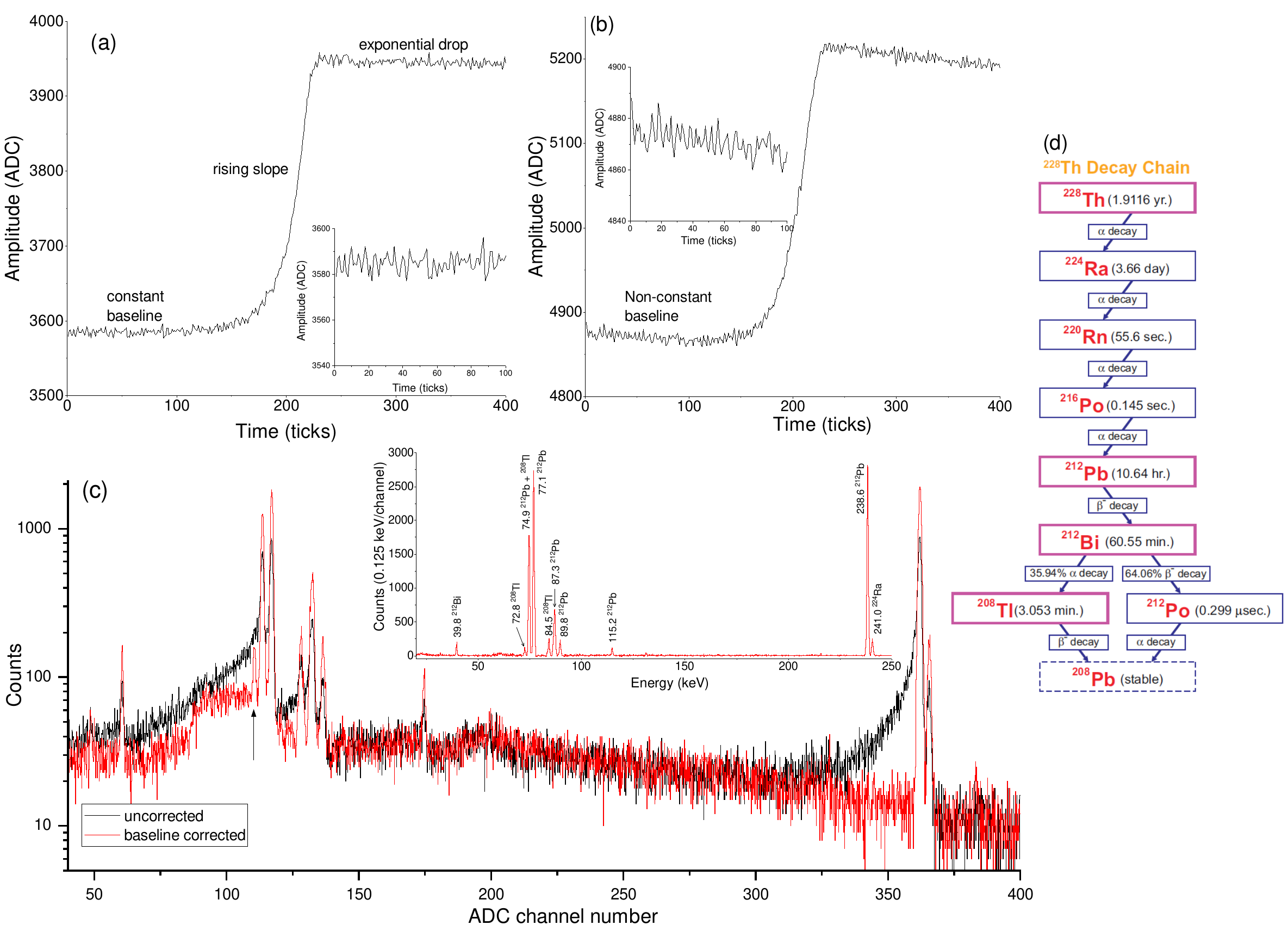}
    \caption[Baseline]{(a) Typical digitized preamplifier pulse from an HPGe detector showing the constant baseline, rising slope and the exponential drop (1 tick = 4 ns). The inset in this figure shows the zoomed version of the same preamplifier pulse to emphasize the constant baseline. (b) Pile-up effects leading to a non-constant baseline of the preamplifier pulse, which is emphasized in the inset of this figure. (c) The uncorrected ADC spectrum (black) and the baseline corrected ADC spectrum (red) for a single HPGe detector for a high activity radioactive $^{228}$Th source. The black arrow indicates the \enquote{hidden} peak (in red) inside the tails (in black). The inset shows the energy calibrated baseline corrected spectrum. (d) Decay chain of $^{228}$Th. 
    }
    \label{fig:baseline}
\end{figure}

A typical HPGe preamplifier pulse/waveform, corresponding to a single muon event, is shown in Fig.~\ref{fig:baseline}(a), with the three different parts indicated.
An important feature to be noted is the constant baseline of the preamplifier pulse, as shown in the inset of Fig.~\ref{fig:baseline}(a).
The maximum amplitude of this pulse is proportional to the energy of the muonic X-ray for that particular muon event. 
At high muon momentum, the muon rate increases (see Fig.~\ref{fig:rates}) and pile-up effects in the HPGe detectors are visible. 
One effect is the partial overlapping of the preamplifier pulses from two consecutive muon events, leading to a non-constant baseline, as shown in Fig.~\ref{fig:baseline}(b) and the inset within. 
This effect leads to an underestimation of the energy deposited by the muonic X-ray in the detector, resulting in a pronounced low-energy tail on the peak in the energy spectra.
By performing baseline correction, the true height of the pulse (\ie the muonic X-ray energy) can be reconstructed. 
These tails are not only seen in in-beam spectra but also for radioactive sources with very high activity.
As an example, the uncorrected ADC spectrum from one single detector, obtained from a highly active $^{228}$Th radioactive source has been shown by the black spectrum in Fig.~\ref{fig:baseline}(c).
As can be see from the red spectrum in Fig.~\ref{fig:baseline}(c), the baseline correction not only removes the low-energy tails but also improves the peak to background ratio.
A notable feature worth mentioning is the peak hidden within the tail marked by the black arrow (see Fig.~\ref{fig:baseline}(c)).
The inset of Fig.~\ref{fig:baseline}(c) shows the energy calibrated baseline corrected spectrum with all the peaks from the different decay products of $^{228}$Th marked (see Fig.~\ref{fig:baseline}(d) for the decay chain of $^{228}$Th)~\cite{nndc}.


\subsubsection{\label{ssubsub:timing_corrections} Timing corrections}
\begin{figure}[ht]
    \centering
    \adjincludegraphics[width=1.0\linewidth, Clip={0\width} {0\height} {0\width} {0\height}]{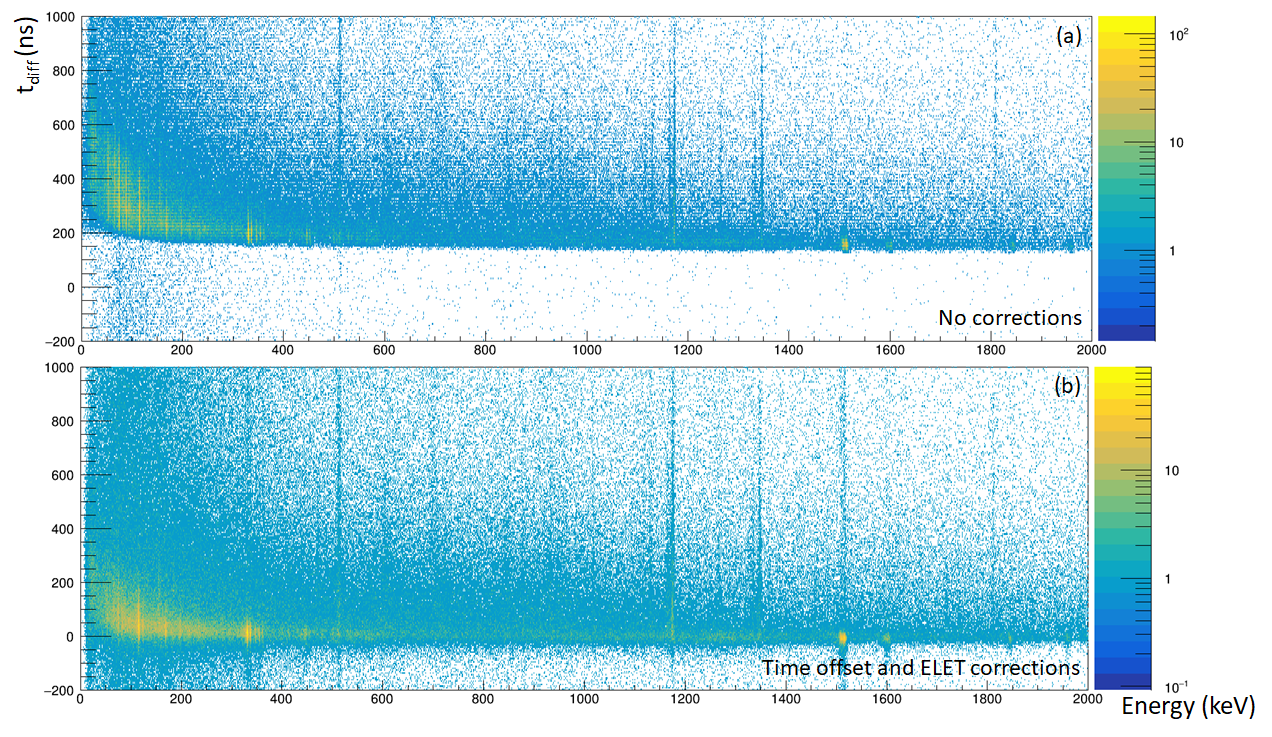}
    \hfill
    \caption[elet]{2D plot of the muonic X-ray energy vs. the $t_{diff}$ for one of the HPGe detectors obtained (a) during the experiment and (b) after the time offset and ELET corrections.}
     \label{fig:elet}
\end{figure}

The SIS3316 digitizer uses the Leading Edge Threshold (LET) algorithm to extract the time from the preampifier pulse, which is unfortunately very sensitive to jitter and walk effects.
Hence in the current analysis, the Extrapolated Leading Edge Threshold (ELET) algorithm has been used in the offline analysis to achieve better timing information.
The measured time $t_{diff}$ of a muonic X-ray event is determined with respect to the observed time of a muon detected in the muon entrance detector.
The 2D plot of the muonic X-ray energy \vs the measured time for one of the HPGe detectors, obtained during the experiment, is shown in Fig.~\ref{fig:elet}(a). 
The time offset and ELET corrections ensure that all the detectors are properly time aligned and that the time resolution is reasonable at all energies (see Fig.~\ref{fig:elet}(b)). 
The best timing resolution we could achieve is $\sim$20~ns at 1~MeV.
Such good timing resolution of detectors is necessary to discriminate between background, prompt and delayed signals.

\subsection{\label{ssub:elements}Elemental muonic X-ray and gamma analysis}
Once all the calibrations and the offline corrections are performed, the spectra 
can be analyzed to quantitatively determine the elemental compositions of the investigated samples. 
For this step, one makes use of existing databases and/or software to determine the muonic X-ray energies and compare it with the experimental spectra to deduce the elements present in the sample~\cite{muxrays_jinr, 2021_Sturniolo}.
To determine the elemental composition,  two different approaches can be followed: (i) make use of calibration curves (\ie ratio of $K_{\alpha}$ or $L_{\alpha}$ lines \vs ratio of at\% of two different elements) obtained from reference standards, which have a similar elemental composition to the investigated sample, or (ii) to directly determine the at\% from the intensities of $K_{\alpha}$ muonic X-ray peaks of the different elements present in the sample~\cite{2022_Biswas_Megatli}. 
For the later approach great care must be taken to correct the observed intensities for the effects of the detector efficiency, the muon capture probability and the branching ratio of the $K_{\alpha}$ peak. 

In both the above approaches, one would need to take into account the mass-attenuation of the muonic X-rays inside the sample itself (see Eq.~\ref{eq:attenuation}). 
For the mass-attenuation effects, there are many added complications which needs to be taken into account for an accurate analysis, like: (i) beam-spot size on the sample, (ii) distribution of muons across the beam-spot, (iii) scattering of muons with the sample,  (iv) contribution of attenuation due to additional collimators/beamdumps, (v) non-uniform density of the sample, (vi) porosity of the sample, (vii) complicated geometry of the sample, (viii) position of the individual detectors on the array, etc.
Since the attenuation coefficient is dependent on the density of the sample, which may not be known for many of the samples, one needs to perform an iterative analysis to arrive at the precise composition. 
This calls for an extensive dedicated Geant4 simulation framework, that is currently under development.

\begin{figure}[ht]
    \centering
    \adjincludegraphics[width=0.9\linewidth, Clip={0\width} {0\height} {0\width} {0\height}]{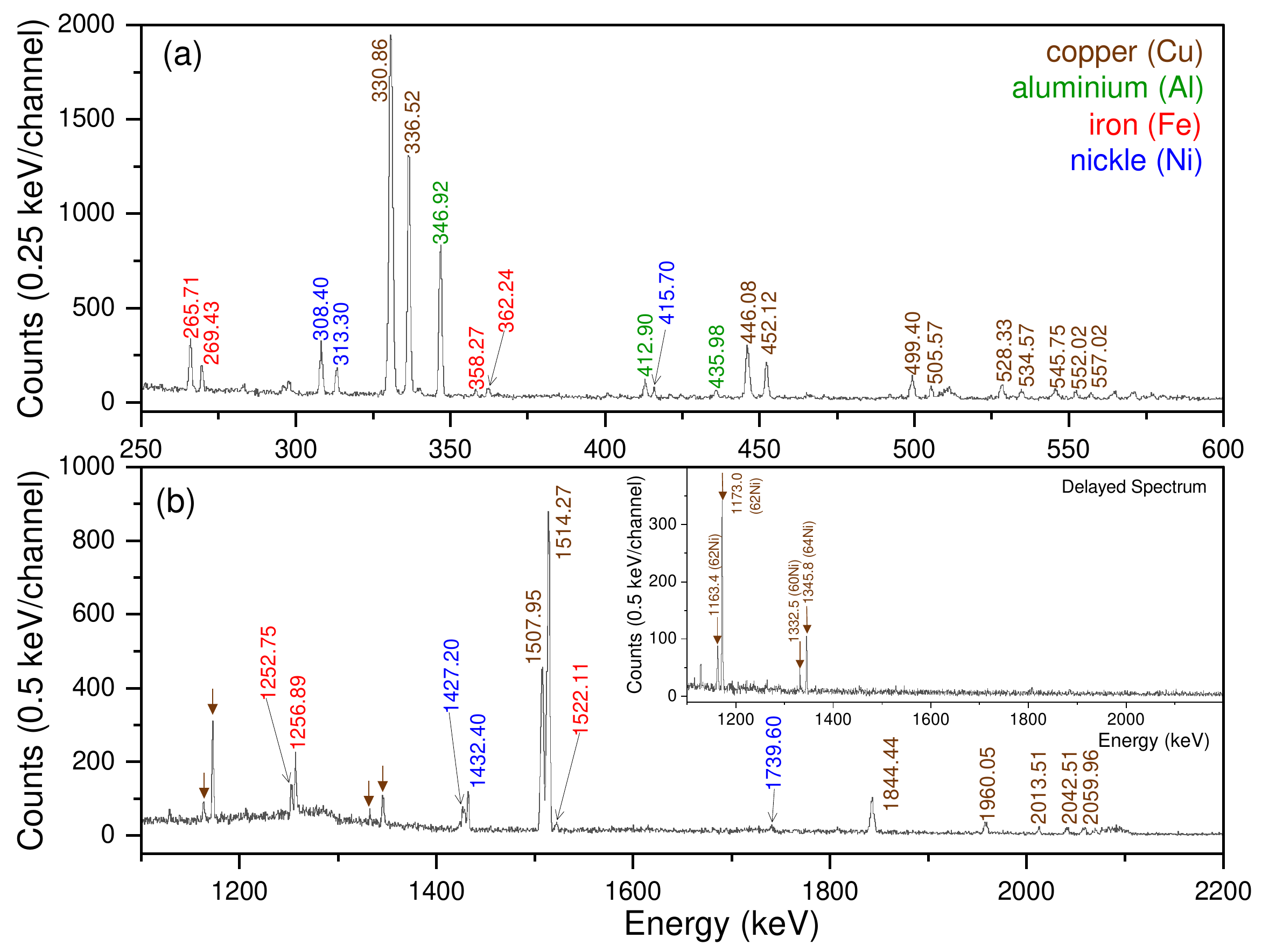}
    \caption[element]{Promt muonic X-ray energy spectrum for the B30 aluminium bronze alloy in the energy ranges (a) 250-600 keV and (b) 1100-2200 keV. The muonic X-ray energies of Cu, Al, Fe, and Ni are shown in brown, green, red, and blue colors, respectively. The inset in (b) shows the delayed spectrum for the same energy range, with the gamma-rays of Ni isotopes marked. 
    }
    \label{fig:element}
\end{figure}

As mentioned in Sec.~\ref{ssubsub:timing_corrections} and in Ref.~\cite{biswas2022}, time-cuts can be performed to separate the prompt and delayed events. 
Figures~\ref{fig:element}(a) and (b) show the prompt spectra ($-50<t_{diff}<50$~ns), from a single HPGe detector, obtained from a B30 aluminium bronze alloy from Schmelzmetall AG~\cite{B30_website} in the energy ranges 250-600~keV and 1100-2200~keV, respectively.
The peaks from the elements Cu, Al, Fe, and Ni can be identified, shown by brown, green, red, and blue colors, respectively.
According to the company's datasheet~\cite{B30_website}, the guaranteed ranges of the chemical composition (in wt\%) for this sample is: 10.5-12.5\% Al, 5.0-7.0\% Fe, 5.0-7.0\% Ni, max. 1.5\% Mn, others 0.5\% and rest is Cu.
The inset in Fig.~\ref{fig:element}(b) shows the delayed spectrum ($100<t_{diff}<200$~ns), where the muonic X-rays are no longer observed and only the gamma-rays from the different isotopes of Ni (nuclear capture of muon by Cu nucleus results in Ni nucleus), are seen.
The gamma-rays are marked with brown arrows in Fig.~\ref{fig:element}(b) and the inset.
The gamma-rays thus act as an additional confirmation to the muonic X-rays on the most abundant element in the sample.

\subsection{\label{ssub:isotopes}Isotope Analysis}
In addition to elemental analysis, the MIXE technique also allows one to identify the isotopic ratios of elements.
Besides the other advantages mentioned above, this is an additional pro over the usually used non-destructive X-ray fluorescence technique (XRF), which cannot identify the isotopes.
The techniques used nowadays for isotopic analysis are Multiple-Collector Inductively Coupled Plasma Mass Spectrometry (MC-ICP-MS), Thermal Ionization Mass Spectrometry (TIMS), and Laser Ablation Inductively Coupled Plasma Mass Spectrometry (LA-ICP-MS), all of which are destructive methods, but provide highly precise isotopic ratios~\cite{2022_Penanes, 2003_Huber, 2003_Boulyga}.

\begin{figure}[ht]
    \centering
    \adjincludegraphics[width=0.6\linewidth, Clip={0\width} {0\height} {0\width} {0\height}]{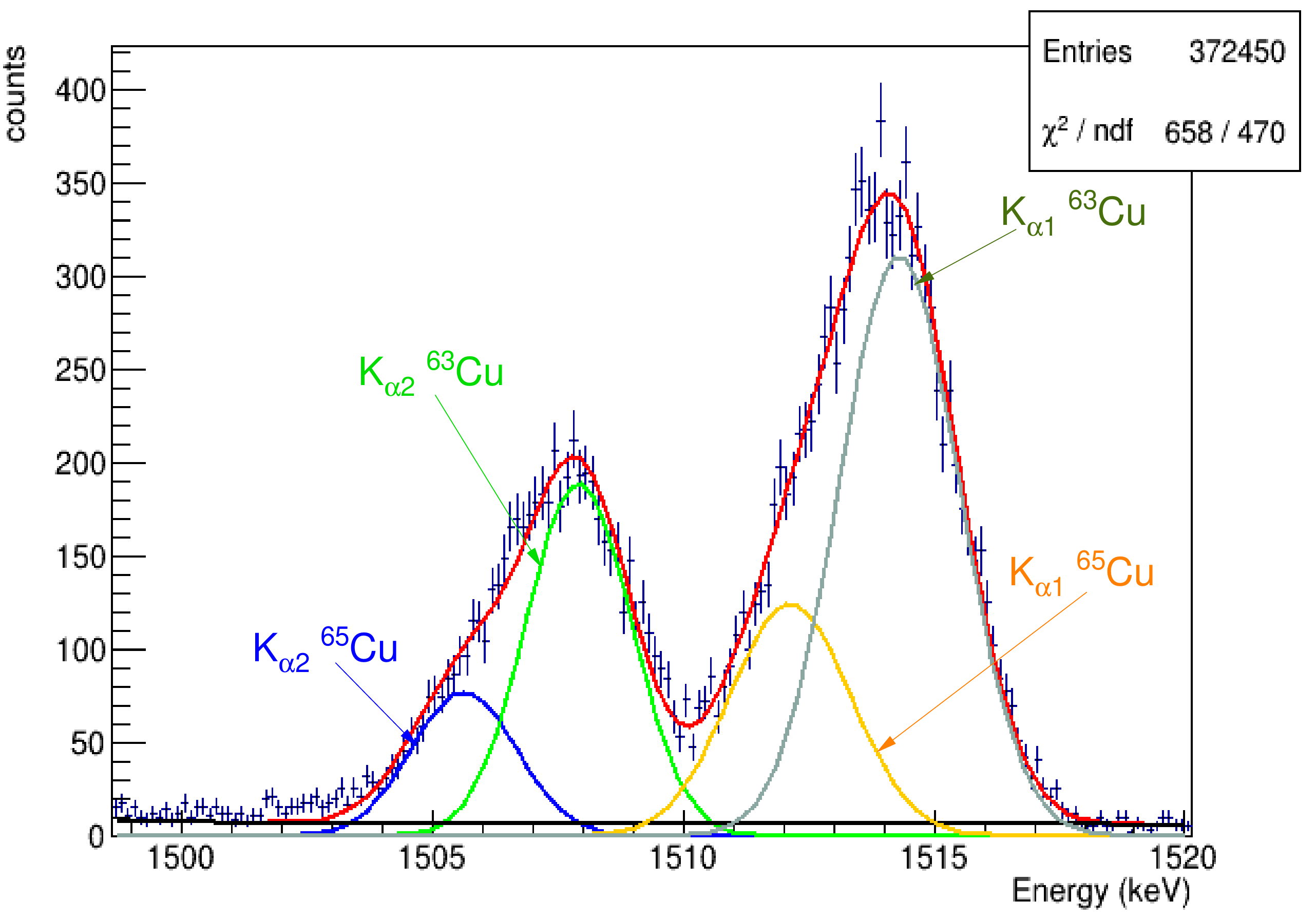}
    \caption[isotope]{Zoomed spectrum in the energy range 1499 to 1520 keV, along with fits, showing the $K_{\alpha1}$ and $K_{\alpha2}$ peaks of the two stable isotopes of Cu, obtained from the MIXE measurement on a pure Cu sample.}
    \label{fig:isotope}
\end{figure}

With the GIANT setup, it has now been possible to determine the isotopic ratios of elements with atomic number $Z > 20$ by using the $K_{\alpha}$ muonic X-ray peak.
An example of such a spectrum, obtained from an individual HPGe detector, for a pure Cu sample, 
along with the fits, is reported in Fig.~\ref{fig:isotope}.
Natural-Cu has two stable isotopes $^{63}$Cu and $^{65}$Cu with natural abundances 69.15(15)\% and 30.85(15)\%, respectively~\cite{nndc}. 
This corresponds to an 
isotopic ratio $^{65}$Cu/$^{63}$Cu = 0.446(2).
The $K_{\alpha1}$ and $K_{\alpha2}$ muonic X-ray energies for these isotopes were determined using the Mudirac software~\cite{2021_Sturniolo}, as shown in the third column of Table~\ref{tab:isotope_cu}.
The fourth column shows that the ratio $K_{\alpha2}/K_{\alpha1} \sim$0.5 for both the Cu isotopes, which is the expected value from theoretical considerations. 
The isotopic ratio obtained for both $K_{\alpha1}$ and $K_{\alpha2}$ transitions, shown in the fifth column, is $\sim$0.4, which is in reasonable agreement with the experimental value of 0.446(2).
However for the proper characterization of the GIANT setup for these isotopes, it is necessary to measure industrial standards and also compare the ratios obtained with samples measured with the other destructive methods.
In addition, the ability to disentangle between muonic X-rays and gamma-rays opens up an alternative route to use gamma-rays (see Section~\ref{ssub:elements}) to determine the isotopic ratios. 

\begin{table}[ht]
    \centering
    \caption[Cu isotopes]{\label{tab:isotope_cu}Muonic X-ray energies and the deduced isotopic ratio for a pure Cu sample}
    \begin{tabular}{lllll}
        \toprule
        Isotope & transition & muonic X-ray energy (keV) & $K_{\alpha2}/K_{\alpha1}$ & $^{65}$Cu/$^{63}Cu$ \\ 
        \midrule
        $^{63}$Cu & $K_{\alpha1}$ & 1514.27 &  &  \\
                  & $K_{\alpha2}$ & 1507.95 & 0.53(2) & \\
        $^{65}$Cu & $K_{\alpha1}$ & 1512.37 &  & 0.40(3)  \\
                  & $K_{\alpha2}$ & 1506.05 & 0.54(4) & 0.41(3)\\
        \bottomrule
    \end{tabular}
\end{table}

An important application of the isotopic ratio determination is in the field of archaeology to deduce the provenance of an excavated artefact.
An important element used in such provenance studies is to determine the isotopic ratios of Pb (which has four stable isotopes $^{204,206,207,208}$Pb) present in the artefact ~\cite{2009_Villa}.
The natural abundance of the these isotopes are $^{208}$Pb(52.4(1)\%), $^{206}$Pb(24.1(1)\%), $^{207}$Pb(22.1(1)\%) and $^{204}$Pb(1.4\%)~\cite{nndc}.
However, since the $K_{\alpha}$ muonic X-ray energy of Pb isotopes is at $\sim$6 MeV, where the efficiencies of the HPGe detectors is low and since the Pb content in such artefacts is also low (usually less than 5\%), this calls for long measurement times in order to determine precise ratios.

\section{Field of applications}
From the above sections, it is evident that the usage of MIXE with \gls{mexp} using the continuous muon beam at PSI is most suitable for doing applied research.
The areas of applied physics, where MIXE has already been used in the campaigns so far, are: (i) Li-ion batteries from Empa, Switzerland~\cite{2022_Arndt} (ii) metallic archaeological artefacts from the Augusta Raurica museum, Switzerland~\cite{2022_Biswas_Megatli} and (iii) meteorites from the Natural History Museum, Bern~\cite{2022_Beda}.

\section{Future Improvements}
The MIXE activities at PSI are ongoing with several projects in the pipeline.
A collaborative effort for an AI based automatized muon beam tuning to rapidly change between momentum settings at optimal conditions has started.
To further optimize the utilization and reduce the workload for users, an automatic sample changer is being planned.
This empowers recent collaborations with industry via ANAXAM (Non-profit technology transfer and access point of large scale research infrastructure at PSI to industry).
Moreover, work on a full scale Geant4 model of the setup and MIXE physics is underway to allow for a rigorous systematic study and higher level of corrections and optimization of MIXE.

\section{Conclusions}
The \gls{mexp} setup was built at PSI and successfully used in multiple MIXE campaigns acquiring up to 500~spectra per year with about three weeks of beamtime.
With a strong focus on versatility and reproducibilty it is forming the basis of the ongoing developments and improvements at LMU.
A single spectra with decent statistics to reach sensitivity of $\approx\SI{1}{\atp}$ is possible within about \SI{1}{\hour} of data taking. 
This is thanks to the high rate but low pile-up of the continuous muon beam at the \gls{smus} (compared to pulsed beams).
The calibration with a set of radioactive sources results in an average detector photo peak efficiency $\overline{\epsilon} \approx \SI{0.11}{\percent}$ and energy resolution $\sigma_E = \SI{0.8}{\keV}$ at $E=\SI{1000}{\keV}$.
During the last campaign (09.2022), a mix of planar and coaxial, totally 14, detectors were used.
Initial corrections using digitized waveforms of the rising edge of the detector signal already proved to be valuable.
For example, isotope fractions of elements like copper with overlapping lines, \ie isotope shift smaller than the energy resolution, can be readily resolved.
The availability of the MIXE technique at PSI has led to a rich user base engaged in various lines of research.
Among already measured samples are archaeological objects, several series of Li-ion battery studies, meteorites, oxidation states of metals, and environmental samples showing unprecedented results.

\section{Acknowledgments}
Our research was funded by the Swiss National Science Foundation, Sinergia project ``Deep$\upmu$'', Grant: 193691 (\url{https://www.psi.ch/en/smus/muon-induced-X-ray-emission-mixe-project}, accessed on 30 August 2022).
The realization of the MIXE project own thanks to numerous people at PSI as also external.
Initial concepts and design choices for the structure profited from discussions with the technical staff of the LMU Hans-Peter Weber and Matthias Elender.
Enormous effort for the operation of the setup during the beamtimes was done by numerous people from the LMU; Zurab Guguchia, Charles Hillis Mielke III, Debarchan Das, Fabian Hotz, Zaher Salman, Toni Shiroka and Chennan Wang. 
Putting the setup together and getting necessary customs parts would have not been possible without the great infrastructure at PSI, \ie the Workshop, Lager/Shop, Liquid Gases Group, Electricians and the  Hallendienst with a special shoutout to Roger Schwarz.
A huge thanks goes to Andrea Raselli, Andreas Suter and Robert Scheuermann for the continued IT, DAQ and MIDAS support.
In general, the collaboration with the muX experiment allowed to kick-start the MIXE endeavour at PSI.
A special thanks within the muX team goes to Frederik Wauters for his nuanced help with the software side of the DAQ.
The mutual assistance with the muMASS experiment for the micromegas and the OMC4DBD collaboration, Osaka university, Tokyo University and RIKEN Nishina Center for Accelerator-Based Science and the Radiochemistry Group of PSI for the mutual lending of HPGes.
Moreover, no research at the HIPA facilities would work without the PSI Division GFA, a tremendous acknowledgement to all personnel involved in the operation of the main proton beam and the secondary beamlines with special mention of Thomas Rauber and Patrick Simon for on-site support at the $\pi$E1 beamline.
Last but not least, technical and physics discussions with Aldo Antognini, Paolo Crivelli, and Anna Soter helped to shape the \gls{mexp} setup.

\printbibliography

\end{document}